\documentclass[aps,preprint,amsmath,amssymb]{revtex4}
\usepackage{graphicx}% Include figure files

\begin{document}

%\title{ \Large\bf $f_{0}(980)$ production in $B$ decay to charmful mesons  }% Force line breaks with \\

\title{  Lepton angular asymmetries
in  semileptonic  charmful B decays }% Force line breaks with \\

\author{ \bf Chuan-Hung Chen$^{a}$\footnote{Email:
physchen@mail.ncku.edu.tw} and Chao-Qiang
Geng$^{b}$\footnote{Email: geng@phys.nthu.edu.tw} }

\affiliation{ $^{a}$Department of Physics, National Cheng-Kung
University, Tainan 701, Taiwan \\
$^{b}$Department of Physics, National Tsing-Hua University,
Hsin-Chu 300, Taiwan }

\date{\today}% It is always \today, today,

\begin{abstract}
We study the lepton angular distributions in $B\to D^{(*)} \ell
\nu_{\ell}$ decays. The lepton angular asymmetries in the
decays with the general effective interactions are examined. We
demonstrate that the asymmetries are sensitive to new physics with
a right-handed quark current.

\end{abstract}

\maketitle
%%%%%%%%%%%%%%%%%%%%%
Although the standard model (SM) has been accepted to be a good
description of physics below the Fermi scale, it is believed that
there is a more fundamental theory at a higher energy scale. Such
theory will generate low energy effective couplings to various
processes, which may differ from those in the SM. To find out the
differences, one needs to search for physical observables which
are sensitive to the new physics.

We shall concentrate on those observables in the semileptonic
charmful $B$ decays $B^-\to D^{(*)}\ell\bar{\nu}_{\ell}$ due to
the large number of $B$'s at the $B$ factories. New physics
effects related to the inclusive decays of $b\to c\ell\nu_{\ell}$
have been studied extensively in the literature
\cite{NP1,NP2,SC,RC,WKN}. In particular, the exclusive decays
$B^-\to D^{(*)}\ell\bar{\nu}_{\ell}$ were used to constrain the
scalar interactions \cite{SC} as well as the vector and
axial-vector interactions \cite{RC} beyond the SM. Moreover, T
violating polarization asymmetries in these exclusive modes were
studied \cite{WKN} in terms of all possible new interactions. In
this report, we examine some asymmetrical physical observables
related to the lepton angular distributions in $B^-\to
D^{(*)}\ell\bar{\nu}_{\ell}$ with the general effective
interactions.

%%%%%%%%%%%%%%%%%%%%%%%%%%%%%%%%%%%%%%%%%%%%
To include the new physics effects, we start with
 the generalized effective Lagrangian for the process
$b\to c\ell\nu_{\ell}$ as
%semileptonic $B$ decays as
\begin{eqnarray}
{\cal L}_{\rm eff} & = & \frac{G_{F}V_{cb}}{\sqrt{2}}\Big\{-
\overline{c} \gamma_{\alpha}(1-\gamma_5)  b
         \overline{\ell} \gamma^{\alpha} (1- \gamma_5) \nu_{\ell}+
     G_V \overline{c} \gamma_{\alpha}  b
         \overline{\ell} \gamma^{\alpha} (1- \gamma_5) \nu_{\ell}
         \nonumber \\&&
  + G_A \overline{c} \gamma_{\alpha} \gamma_5 b
         \overline{\ell} \gamma^{\alpha} (1- \gamma_5) \nu_{\ell}+  G_S
\overline{c} b \overline{\ell} (1 - \gamma_5) \nu_{\ell}
   + G_P \overline{c} \gamma_5 b  \overline{\ell} (1 - \gamma_5)
    \nu_{\ell}+ {\rm h.c.}\Big\} , \label{leff}
\end{eqnarray}
where $G_F$ is the Fermi constant, $V_{cb}$ is the relevant CKM
matrix element, and
%%, $G^{\prime}_{V(A)}=G_{V(A)}\mp 1$.
$G_i\ (i=S,P,V,A)$ denote the strengths of the new
effective scalar, pseudoscalar, vector and axial-vector
interactions, respectively. The tiny contributions from right
handed neutrino are neglected. We note that, in general,
the tensor interactions,
described by $\bar{c}\sigma_{\mu\nu}(1\pm\gamma_5)b
\bar{\ell}\sigma^{\mu\nu}(1-\gamma_{5})\nu_{\ell }$ with
$\sigma_{\mu\nu}=i[\gamma_{\mu},\gamma_{\nu}]/2$, should be also included
in Eq.~(\ref{leff}).
However,
since these tensor interactions
 usually arise from the models associated with
baryon and lepton number violations, such as leptoqaurk
models \cite{LQ}, for simplicity, we will not consider their effects.
Instead, we will
only concentrate on the
models with baryon and lepton number conservations, such as
the minimal supersymmetric standard model (MSSM).
% Hence, we neglect to consider the effects of tensor interactions.

To study the exclusive semileptonic $B$ decays, such as $B^-\to
D^{(*)0}\ell^-\bar{\nu}_{\ell}$, in terms of ${\cal L}$ in Eq.
(\ref{leff}), we need to know the form factors in the transition
matrix elements $\langle D^{(*)}| \bar{c} \Gamma b|B\rangle$.
%transition form factors for the exclusive semileptonic B decays,
We parameterize the relevant transition matrix elements to be
\cite{WKN},
\begin{eqnarray}
\langle D(p^{\prime})| \overline{c} \gamma_{\mu} b |B(p)
\rangle
 & = & f_+(q^{2}) \, (p+p^{\prime})_{\mu}
      + f_-(q^2) \, (p-p^{\prime})_{\mu}\,;
\label{FF1} \\
\nonumber\\
\langle D^*(p^{\prime}, \epsilon)| \overline{c} \gamma^{\mu} b
|B(p) \rangle
 & = &  i \frac{F_V(q^2)}{m_B} \epsilon^{\mu\nu\alpha\beta}
         \epsilon^*_{\nu}  (p+ p^{\prime})_{\alpha} q_{\beta}\,,
          \nonumber \\
\langle D^*(p^{\prime}, \epsilon)| \overline{c} \gamma_{\mu}
   \gamma_5 b |B(p) \rangle
 & = & - F_{A0}(q^2) \, m_B \epsilon^*_{\mu}
 -\frac{F_{A1}(q^2)}{m_B} (p+p^{\prime})_{\mu} \epsilon^* \cdot q
 -\frac{F_{A2}(q^2)}{m_B} q_{\mu} \epsilon^* \cdot q \, ,
 \label{FF2}
\end{eqnarray}
where $p$ and $p^{\prime}$ are the four-momenta of $B$ and $D$
($D^*$), respectively, $\epsilon$ is the polarization vector of
$D^*$ meson, and $q=p-p^{\prime}$. By
using equation of motion, the hadronic matrix elements for scalar
and pseudoscalar currents are given by
\begin{eqnarray}
\langle D(p^{\prime})| \overline{c} b |B(p) \rangle
& = & \frac{m_B^2}{m_b-m_c}\left[ f_+(q^2)  (1-r_D) + f_-(q^2) \frac{q^2}{m_B^2}\right], \\
%\langle D(p^{\prime})| \overline{c} \gamma_5 b |B(p) \rangle
%& =& 0 \\
%\langle D^*(p^{\prime}, \epsilon)| \overline{c} b |B(p) \rangle
%&=&0 \\
\langle D^*(p^{\prime}, \epsilon)| \overline{c} \gamma_5 b
|B(p) \rangle & =& \frac{m_B}{m_b+m_c} \epsilon^* \cdot q
    \left[ F_{A0}(q^2) + F_{A1}(q^2) (1-r_{D^*}) + F_{A2}(q^2)  \frac{q^2}{m_B^2}\right],
\label{FF5}
\end{eqnarray}
where $m_{b,c}$ are the quark masses and
$r_{D^{(*)}}=m_{D^{(*)}}^2/m_B^2$.
We note that the matrix elements  $\langle D(p^{\prime})| \overline{c}
\gamma_{\mu} \gamma_5 b |B(p)\rangle$, $\langle D(p^{\prime})|
\overline{c} \gamma_5 b |B(p) \rangle$ and $\langle
D^*(p^{\prime}, \epsilon)| \overline{c} b |B(p) \rangle$ are equal
to zero due to parity and helicity since
  $B\ (D)$ and $D^*$
are pseudoscalar and vector mesons, respectively.

 From the interactions in Eq. (\ref{leff}) and the
 form factors in Eqs. (\ref{FF1})-(\ref{FF5}),
 the decay amplitudes  can be written as
\begin{eqnarray}
{\cal A}_{D}=\langle D\ell \nu_{\ell}| {\cal
L}_{eff}|\bar{B}\rangle=\sigma \bar{\ell} (1-\gamma_5)\nu_{\ell}+
j^{\mu} \bar{\ell} \gamma_{\mu}(1-\gamma_5) \nu_{\ell}
\end{eqnarray}
for $B\to D \ell
\nu_{\ell}$, where $\sigma=G_{S}\langle D|\bar{c} b|\bar{B}\rangle$ and
$j^{\mu}=G^{\prime}_{V}\langle D|\bar{c}
\gamma^{\mu}b|\bar{B}\rangle$, and
 \begin{eqnarray}
{\cal A}_{D^*}=\langle D^*(\epsilon)\ell \nu_{\ell}| {\cal
L}_{eff}|\bar{B}\rangle=\Sigma(\epsilon) \bar{\ell} (1-\gamma_5)
\nu_{\ell}+ J^{\mu}(\epsilon) \bar{\ell} \gamma_{\mu} (1-\gamma_5)
\nu_{\ell}
\end{eqnarray}
for $B\to D^* \ell \nu_{\ell}$, where $\Sigma(\epsilon)=G_{P}\langle
D^*(\epsilon)|\bar{c}\gamma_5 b|\bar{B}\rangle$ and
$J^{\mu}(\epsilon)=G^{\prime}_{V}\langle D^*(\epsilon)|\bar{c}
\gamma^{\mu}b|\bar{B}\rangle + G^{\prime}_{A} \langle
D^*(\epsilon)|\bar{c} \gamma^{\mu}\gamma_5 b|\bar{B}\rangle$ with
$G^{\prime}_{V(A)}=G_{V(A)}\mp 1$. Since our purpose is to study
the lepton angular distributions, we evaluate the
decay amplitudes in the rest frame of the lepton pair invariant
mass $q^2$. The kinematical variables for particles are chosen to
be
\begin{eqnarray}
q&=&(\sqrt{q^2},0,0,0),\ \ \ p=(E_B,0,0,|\vec{p}_X|), \ \ \
p^{\prime}=(E_X,0,0,|\vec{p}_X|), \nonumber \\
 p_{\ell}&=&(E_{\ell},|\vec{p}_{\ell}| \sin\theta, 0,
 |\vec{p}_{\ell}| \cos\theta), \ \ \
 |\vec{p}_X|=\frac{Z(m_{X})}{2\sqrt{q^2}}, \ \ \
 E_{B}=\sqrt{|\vec{p}_X|^2+m^2_B}, \nonumber \\
 E_{X}&=&\sqrt{|\vec{p}_X|^2+m^2_X},\ \ \
 Z(m_{X})=\sqrt{(m^{2}_{B}-(m_X-\sqrt{q^2})^{2})(m^{2}_{B}-(m_X+\sqrt{q^2})^{2})}, \nonumber
 \\
|\vec{p}_{\ell}|&=&(q^2-m^2_{\ell})/(2\sqrt{q^2}),\ \ \
\epsilon(0)=\frac{1}{m_{D^*}}(|\vec{p}_{D^*}|,0,0,E_{D^*}), \ \ \
\epsilon(\pm)=\frac{1}{\sqrt{2}}(0,1,\pm i, 0),
\end{eqnarray}
where $X$ denotes $D$ or $D^*$.
 It is clear that  $\theta$ is defined as the polar angle
of the lepton momentum relative to the moving direction of the $B$-meson
in the $q^2$ rest frame. Hence, based on our conventions, the
differential decay rate with respect to the invariant mass $q^2$
and the lepton polar angle  for $B\to D \ell \nu_{\ell}$ is
described by
\begin{eqnarray}
{d\Gamma \over dq^2 d\cos\theta
}&=&{G^{2}_{F}|V_{cb}|^{2}|\vec{P}_{D1}| |\vec{p}_{\ell}|^{2} \over
2^{6} \pi^{3} m^{2}_{B}}\left(1-\frac{m^{2}_{\ell}}{q^{2}}\right)
%{\cal P}_{D},
\left\{
\frac{\sqrt{q^2}}{|\vec{p}_{\ell}|} \left[
|\rho_D|^{2}+ \frac{m^2_{\ell}}{q^2}|G|^{2} \right]
+\right.
\nonumber\\
%2\frac{m_{\ell}}{|\vec{p}_{\ell}|} {\Re}(\rho_{D}\; G^{*})
&&\left.A_D\cos\theta +2|G|^{2}\sin^{2}\theta \right\}\, ,
\label{diffd}
\end{eqnarray}
%%%%%%%
where
\begin{eqnarray}
A_{D}&=&
2\frac{m_{\ell}}{|\vec{p}_{\ell}|} Re\,(\rho_{D}\; G^{*}),
\nonumber\\
%%%%%%%%%%%%%%%%%%%%%%%%%%%%%%
\rho_D&=&\sigma+\frac{m_{\ell}}{q^2} G^{\prime}_{V} \left(q^2
f_{-}(q^{2})+(m^2_B-m^2_D)f_{+}(q^2)\right),\ \ \
%\sigma=G_{s}\langle D |c\,\bar{b} |\bar{B}\rangle,
\nonumber \\
G&=&2|\vec{p}_{D}|f_{+}(q^{2})G^{\prime}_{V}, \ \ \
|\vec{P}_{D1}|= \frac{Z(m_{D})}{2m_{B}}.
\label{ppd}
\end{eqnarray}
The differential decay rate for $B\to D^{*} \ell \nu_{\ell}$ is
expressed as
\begin{eqnarray}
{d\Gamma \over dq^2
d\cos\theta}&=&{G^{2}_{F}|V_{cb}|^{2}|\vec{P}_{D^*1}||\vec{p}_{\ell}|^{2}
 \over 2^{6} \pi^{3}
m^{2}_{B}}\left(1-\frac{m^{2}_{\ell}}{q^{2}}\right)
%{\cal P}_{D^*},
 \left\{Y_{0}+A_{D^*} \cos\theta+2|J_{3}|^{2}\sin^{2}\theta
\right. \nonumber\\
 &&\left.+\left(|J_{+}|^2+|J_{-}|^{2}\right)\cos^{2}\theta
\right\}\, ,
 \label{diffvd}
\end{eqnarray}
%%%%%%%
where
\begin{eqnarray}
A_{D^*}&=&
\frac{2m_{\ell}}{|\vec{p}_{\ell}|} Re\,(\rho_{D^*} J^*_{3})+
\frac{\sqrt{q^{2}}}{|\vec{p}_{\ell}|}
\left(|J_{+}|^2-|J_{-}|^{2}\right)\, , \nonumber \\
%%%%%%%%%%%%%%%%%%%%%%%%%%%%%%
Y_{0}&=&\frac{\sqrt{q^2}}{|\vec{p}_{\ell}|}\left(|\rho_{D^*}|^{2}
+
\frac{m^{2}_{\ell}}{q^2}|J_{3}|^{2}+\frac{E_{\ell}}{\sqrt{q^2}}\left(|J_{+}|^2+|J_{-}|^2\right)
\right)\, , \nonumber\\
|\vec{P}_{D^*1}| &=& \frac{Z(m_{D^*})}{2m_{B}}\,,
\label{pvd}
\end{eqnarray}
with
\begin{eqnarray}
 \rho_{D^*}&=&\Sigma+\frac{m_{\ell}}{\sqrt{q^2}}J_{0},
\nonumber \\
%%%%%%%%%%%%%%%%%%%%
%%%%%%%%%%%%%%%%%%%%%
%\rho&=&\frac{m_{B}}{m_{b}+m_{c}}G_{P}\left(F_{A0}(q^{2})
%+(1-r_{D^*})F_{A1}(q^2)+\frac{q^2}{m^2_{B}}F_{A2}(q^2)
%\right)-\frac{m_{\ell}}{m_{B}} G^{\prime}_{A} F_{A1}(q^2) \nonumber \\%%%%%%%%%%%%
%%%%%%%%%%%%%%%%%%%%%%%%%%%%%%%%%%%%%%%%%
J_{0}&=&-G^{\prime}_{A} \frac{m_{B}}{m_{D^*}} |\vec{p}_{D^*}|
\left[
F_{A0}(q^2)+F_{A1}(q^2)(1-r_{D^*})+F_{A2}(q^2)\frac{q^2}{m^2_B}\right],
\nonumber \\ %%%%%%%%%%%%%%%%%%%%%,
%%%%%%%%%%%%%%%%%%%%%%%%%%%%%%%%%%%
J_{3}&=&-G^{\prime}_{A} \frac{m_{B}}{m_{D^*}}E_{D^*}
 \left[
F_{A0}(q^2)+2F_{A1}(q^2)\frac{|\vec{p}_{D^*}|^2}{m^{2}_{B}}\frac{\sqrt{q^2}}{E_{D^*}}\right],
\nonumber \\ %%%%%%%%%%%%%%%%%%%%%
J_{\pm}&=& \frac{1}{\sqrt{2}}\left(m_{B}G^{\prime}_{A} F_{A0}(q^2)
\pm 2
\frac{|\vec{p}_{D^*}|}{m_{B}}\sqrt{q^2}G^{\prime}_{V}F_{V}(q^2)\right)\,.
\end{eqnarray}
$\rho_{D^{*}}$, $J_{0}$ and $J_{3}$ denote the longitudinal
contributions of $D^*$, while $J^{\pm}$ are the transverse effects.

Since the differential decay rates in Eqs.~(\ref{diffd}) and
(\ref{diffvd}) involve the polar angle of the lepton, we can define an
angular asymmetry to be
\begin{eqnarray}
{\cal A}(q^2)={\int^{\pi/2}_{0}d\cos\theta
d\Gamma/(dq^2d\cos\theta)-\int^{\pi}_{\pi/2}d\cos\theta
d\Gamma/(dq^2d\cos\theta) \over \int^{\pi/2}_{0}d\cos\theta
d\Gamma/(dq^2d\cos\theta)+\int^{\pi}_{\pi/2}d\cos\theta
d\Gamma/(dq^2d\cos\theta)},\label{asy}
\end{eqnarray}
 from which we may study the behavior of the lepton angular
distributions in Eqs. (\ref{diffd}) and (\ref{diffvd}). It is easy
to see that the asymmetry in Eq. (\ref{asy}) is related to the
parity-odd terms associated with $\cos\theta=\vec{p}_{X}\cdot
\vec{p}_{\ell}/|\vec{p}_{X}||\vec{p}_{\ell}|$, appearing in the
formulas of the differential decay rates in Eqs. (\ref{diffd}) and
(\ref{diffvd}). Explicitly, we have
\begin{eqnarray}
{\cal A}(q^2) \propto A_{D^{(*)}}
\label{asyf}
\end{eqnarray}
for $B\to D^{(*)}\ell\nu_{\ell}$, where $A_{D^{(*)}}$ are defined
in Eqs. (\ref{ppd}) and (\ref{pvd}), respectively. For $B\to D
\ell \bar{\nu}_{\ell}$ ($\ell=e,\mu,\tau$) decays, the parity-odd
effects could be only generated by the interference between scalar
and vector interactions. To fit the proper chirality, therefore,
we need one lepton mass insertion. That is the reason why
$A_D$ in Eq.~(\ref{ppd}) is proportional to the
lepton mass. Thus, the asymmetries in  the electron and muon modes
are
negligible, but it could be large in the $\tau$ mode. However,
since $D^{*}$ carries the transverse degree of freedom, the parity-odd
effect could be induced from such extra degree, even
in the chiral limit of $m_{\ell}=0$. From Eq.~(\ref{FF2}), we see
that the transverse effect is related to form factors $F_{V}$
and $F_{A0}$. One expects that $A_{D^*}$ will be proportional
to $G^{\prime}_{V}F_V G^{\prime}_{A} F_{A0}$. Hence, According to
Eq.~(\ref{pvd}), besides the term proportional to the lepton
mass, in the $D^*$ production mode we have the contribution from
$|J_+|^2-|J_{-}|^2\propto Re\,( G^{\prime}_V
G^{\prime*}_{A})F_{V}F_{A0} $ due to the transverse
polarization of $D^*$ \cite{TP}. It is worth mentioning that if
the couplings involved are symmetric in parity, {\it i.e.},
$G^{\prime}_{V}=0$ or $G^{\prime}_{A}=0$, we still cannot get
the angular asymmetry in $B\to D^{*} \ell \bar{\nu}_{\ell L}$ in
the chiral limit.

\begin{figure}[htbp]
%\includegraphics*[width=1.6
%in]{one-loop}
\includegraphics*[width=2.3in]{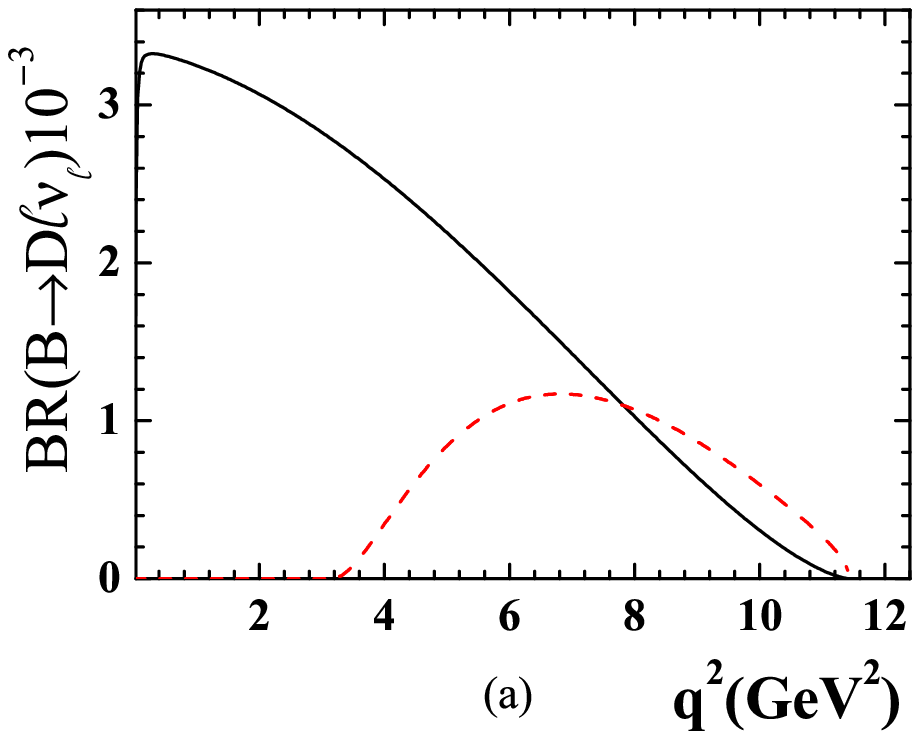} \hspace{0.3cm} \includegraphics*[width=2.4in]{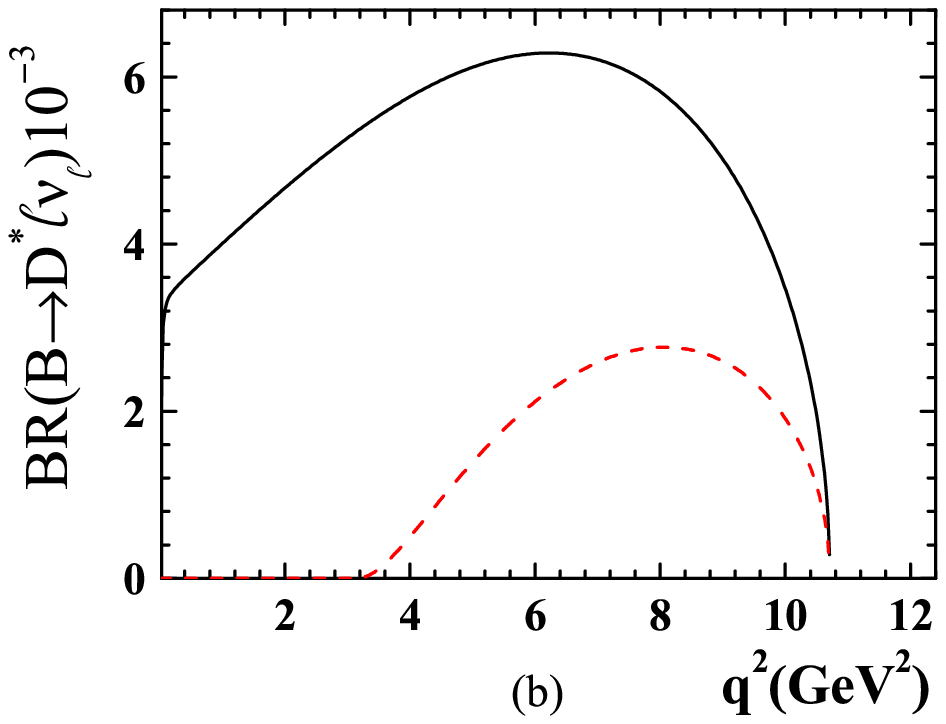}\caption{
Differential decay rates in the SM for (a) $B^{-}\to D^{0}
\ell^{-} \bar{\nu}_{\ell}$ and (b) $B\to D^{*0} \ell^{-}
\bar{\nu}_{\ell}$. The solid lines denote $\ell=e$ and $\mu$, while
the dashed lines are $\ell=\tau$.}
 \label{figrate_sm}
\end{figure}

\begin{figure}[htbp]
%\includegraphics*[width=1.6
%in]{one-loop}
\includegraphics*[width=2.3in]{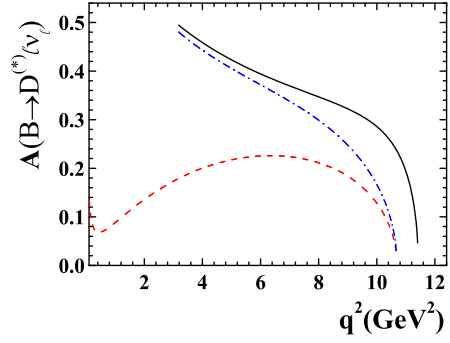} \caption{
Angular asymmetries in the SM for $B^{-}\to D^{0} \tau^{-}
\bar{\nu}_{\tau}$ (solid) , $B^{-}\to D^{*0} \ell^{-}
\bar{\nu}_{\ell}$ ($\ell=e$ and $\mu$, dashed) and $B^{-}\to D^{*0} \tau^{-}
\bar{\nu}_{\tau}$ (dash-dotted). }
 \label{figasy}
\end{figure}

To get the numerical values, the transition form factors
based on the heavy quark symmetry are taken to be \cite{WKN}
\begin{eqnarray}
f_{\pm}&=&\pm \frac{1\pm \sqrt{r_D}}{2r_{D}^{1/4}}\xi_{1}(w), \ \ \
F_V=\frac{1}{2r_{D}^{1/4}}\xi_{2}(w),\ \ \
F_{A0}=-r^{1/2}_{D}(1+w)\xi_{2}(w),\nonumber\\
F_V&=&F_{A1}=-F_{A2}\,,
\end{eqnarray}
%and $F_V=F_{A1}=-F_{A2}$.
where $\xi(w)_{i}$ are the Isgur-Wise
functions, which are normalized to unity at zero recoil, and
$w=(m^{2}_{B}+m^{2}_{X}-q^2)/(2m_{B}m_X)$. We note that to
include the correction due to the heavy quark symmetry breaking, we
adopt the Isgur-Wise (IW) functions for $D$ and $D^*$ productions
as $\xi_{1}(w)=1-0.75(w-1)$ and $\xi_2(w)=1-0.95(w-1)$,
respectively, based on the results in Ref. \cite{GL}.
 Following  Eqs.~(\ref{diffd}) and
(\ref{diffvd}), the decay branching ratios (BRs) for $B\to D^{(*)}
\ell \bar{\nu}_{\ell}$ in the SM
are summarized in Table \ref{tablebr}. In the table, we also show
 the current experimental data given by Ref. \cite{PDG}.
 We note that the decay BRs for the light lepton modes of $e$ and $\mu$
 are insensitive to the lepton masses.
\begin{table}[phbt]
\caption{Decay BRs  (in units of $10^{-2}$) of  $B^-\to D^{(*)}
\ell \bar{\nu}_{\ell}$ ($\ell=e,\mu$) and $B^-\to
D^{(*)}\tau\bar{\nu}_{\tau}$.}\label{tablebr}
\begin{ruledtabular}
\begin{tabular}{ccccc}
 Decays & $B^-\to D^0 \ell \bar{\nu}_{\ell}$
  & $B^-\to D^0 \tau \bar{\nu}_{\tau}$ & $B^-\to D^{*0} \ell \bar{\nu}_{\ell}$&
     $B^-\to D^{*0} \tau \bar{\nu}_{\tau}$ \\
  \hline
 SM  & 2.07 & 0.62 & 5.43 & 1.39 \\
\hline
Experiments \cite{PDG} & $2.15\pm 0.22$& & $6.5\pm 0.5$ & \\
  \end{tabular}
\end{ruledtabular}
\end{table}
The differential decay rates
and angular asymmetries
 in the SM
 are displayed in
Figs.~\ref{figrate_sm} and
\ref{figasy}, respectively.
%
 %%%%%%%%%%%%%%%%%%%%%%%%%%%%%%%%%%%%%%%%%%%%%%%%%%%%%%%%%%%%%
 %%%%%%%%%%%%%%%%%%%%%%%%%%%%%%%%%%%%%%%%%%%%%%%%%%%%%%%%%%%%%
 %The angular asymmetries
 %in the SM are presented in Fig.~\ref{figasy}.
 %%%%%%%%%%%%%%%%%%%%%%%%%%%%%%%%%%%%%%%%%%%%%%%%%%%%%%%%%%%%%
 %%%%%%%%%%%%%%%%%%%%%%%%%%%%%%%%%%%%%%%%%%%%%%%%%%%%%%%%%%%%%
 %In the figure, we neglect the small effects on
 % $B\to D \ell \bar{\nu}_{\ell}$ for $\ell=e,\; \mu$.
%%% We note that although the asymmetry of $B\to D
%%% \tau \bar{\nu}_{\tau}$ in the SM does not relate to the parity
%%%  violating terms in quark sector, since only left-handed current
%%% involved in lepton sector, we still can have parity violating
%%% effects by the extra one more lepton mass. The similar effects
%%% also appear in the decays $B\to D^* \ell \bar{\nu}_{\ell}$.

\begin{figure}[htbp]
%\includegraphics*[width=1.6
%in]{one-loop}
\includegraphics*[width=1.8in]{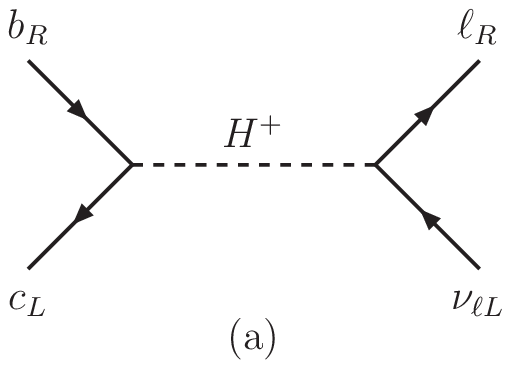}\hspace{0.3cm}  \includegraphics*[width=1.8in]{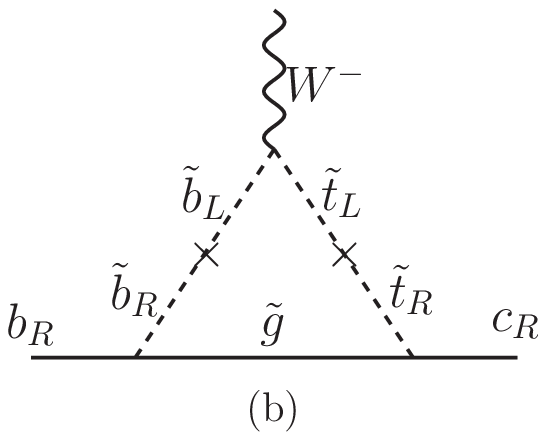}
  \caption{
Feynman diagrams for effective interactions (a) ${\cal L}_H$ and
(b) ${\cal L}_{R}$.}
 \label{feyn}
\end{figure}

To illustrate  new physics effects on  the angular
asymmetries, we consider two types of new interactions. One of them
is the interactions arising from a charged Higgs, described by
\begin{eqnarray}
{\cal L}_H=\frac{G_F}{\sqrt{2}}V_{cb}C^{\ell}_{H} \bar{c}
(1+\gamma_5) b \bar{\ell} (1-\gamma_{5})\nu_{\ell}.
\end{eqnarray}
The other one is due to the right-handed current in the quark sector, given by
\begin{eqnarray}
{\cal L}_R=\frac{G_F}{\sqrt{2}}V_{cb}C_{R} \bar{c}
\gamma_{\mu}(1+\gamma_5) b \bar{\ell} \gamma^{\mu}
(1-\gamma_{5})\nu_{\ell}.
\end{eqnarray}
From Eqs.~(\ref{ppd}) and (\ref{asyf}), we find that the influence of $G_S$
on the $B\to D \tau \bar{\nu}_{\tau}$ decay is much effective than
that of $G_V$. Therefore, we only consider the contribution of
${\cal L}_{H}$ to $B\to D \tau \bar{\nu}_{\tau}$. On the other hand,
as
the transverse polarizations of $D^*$ are sensitive to
the angular asymmetries in $B\to D^* \ell \bar{\nu}_{\ell}$
decays, the influences of $G_{A(V)}$ are much effective than that of
$G_P$. Hence, we only concentrate $G_{A(V)}$ for $B\to D^* \ell
\bar{\nu}_{\ell}$ decays.
%In order to avoid giving too large BRs
%on these semileptonic $B$ decays,
It is clear that the parameters with the new interactions need to
satisfy the current experimental data \cite{PDG} shown in Table
\ref{tablebr}.
%  $BR(B^-\to D^0 \ell \bar{\nu}_{\ell})=(2.15\pm 0.22)\%$
%and $BR(B^-\to D^{*0} \ell\bar{\nu}_{\ell})=(6.5\pm 0.5)\%$.
 For more specific models,
we adopt those governed by supersymmetry with R-parity invariance
as shown in Ref. \cite{WKN}. The Feynman diagrams with the tree
and one-loop corrections to the SM are shown in Fig.~\ref{feyn}.
 %%%%%%%%%%%%%%%%%%%%%%%%%%%%%%%%%%%%%%%%%%%%%%%%%%%%%%%%%%%%%
 %%%%%%%%%%%%%%%%%%%%%%%%%%%%%%%%%%%%%%%%%%%%%%%%%%%%%%%%%%%%%
The corresponding $C^{\ell}_{H}$ and $C_R$ are given by
      \begin{eqnarray}
         C^{\ell}_{H}&=& -\frac{1}{m^{2}_{H}}m_b m_{\ell} \tan^{2}\beta\, ,
         \nonumber \\
         C_{R}&=&-\frac{\alpha_s}{36\pi}\frac{m_t (A_t-\mu\cot\beta)}{m^2_{\tilde{g}}} \frac{m_b(A_b-\mu\tan\beta)}{m^2_{\tilde{g}}}
         \frac{\tilde{V}_{33}V^{U_R*}_{32}V^{D_R}_{33}}{V_{cb}}
         I_{0}\left(\frac{m^{2}_{\tilde{t}}}{m^{2}_{\tilde{g}}}, \frac{m^{2}_{\tilde{b}}}{m^{2}_{\tilde{g}}}\right)\, , \nonumber \\
         I_{0}\left(a,b\right)&=&
         \int^{1}_{0} dz_{1}\int^{1-z_1}_{0}dz_{2}{24 z_{1}z_{2} \over [a z_1 +b z_2
         +(1-z_1-z_2)]^2},\label{nwc}
      \end{eqnarray}
where $A_{t(b)}$ are the soft SUSY breaking $A$ terms, $\mu$
stands for the two Higgs superfields mixing parameter, $\tan\beta$
is the ratio of the two Higgs vacuum expectation values (VEVs),
and $m_i$ ($i=\tilde{g},\tilde{t},\tilde{b},H$) are
the masses of gluino, stop, sbottom and charged Higgs,
respectively. Since squarks and quarks in general have different
flavor structures, the unitary matrices to diagonalize the mass matrices
of  up(down)-quark and those of their superpartners are also
different. For simplicity, we choose the bases that the mass
matrices of squarks and quarks are diagonalized before including
the soft SUSY breaking $A$ terms, which govern the mixings of left
handed and right-handed squarks. The effects of $A$ terms, the 2nd
and 3rd factors of $C_{R}$ in Eq.~(\ref{nwc}), could be taken as
perturbations. We use $\tilde{V}_{33}$ to denote the super
CKM matrix associated with the coupling $W^{-}
 \tilde{b}^*_{L}\tilde{t}_{L}$ while $V^{U_{R}(D_R)}$ are the
 mixing matrices for diagonalizing the upper (down) type quarks.

\begin{figure}[htbp]
%\includegraphics*[width=1.6
%in]{one-loop}
\includegraphics*[width=2.in]{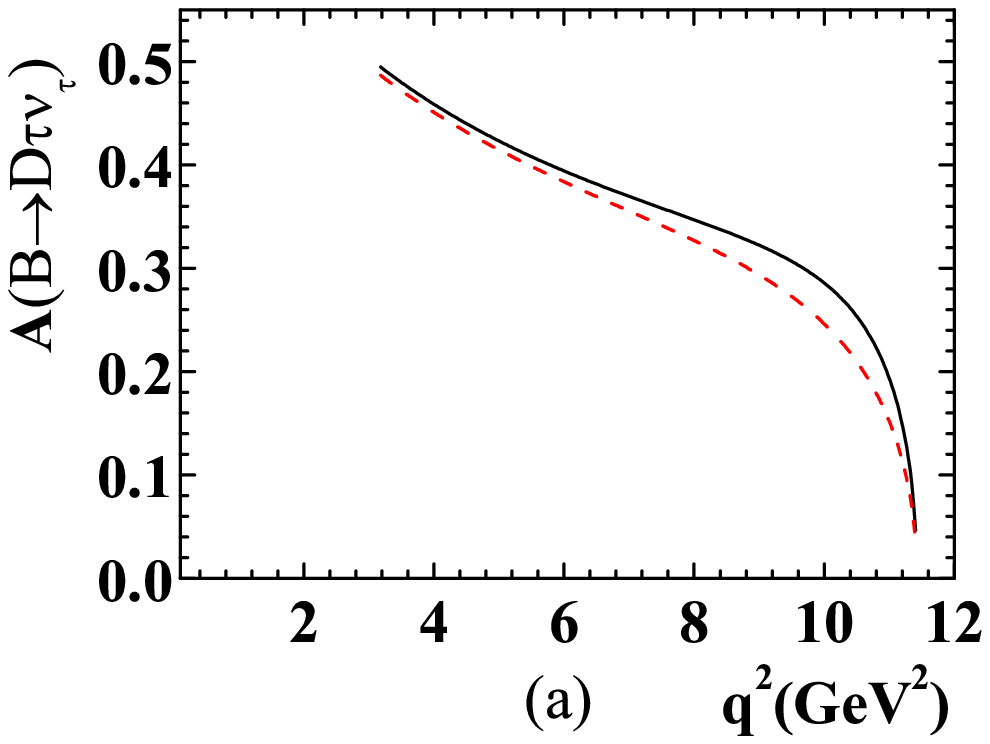}
\includegraphics*[width=2.in]{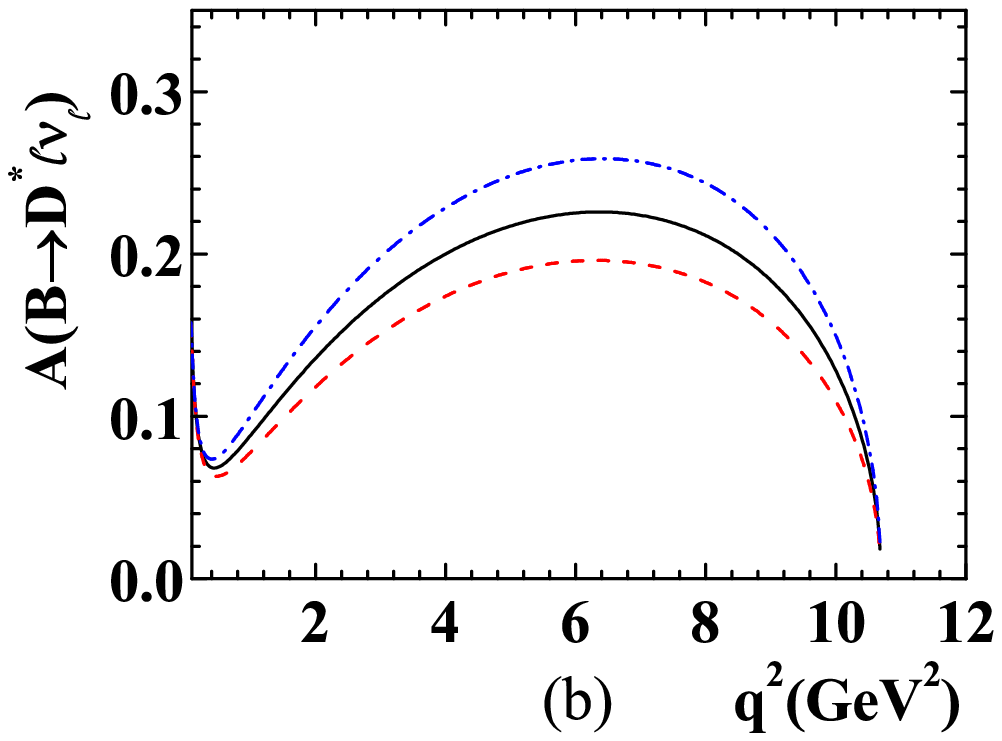}
 \includegraphics*[width=2.in]{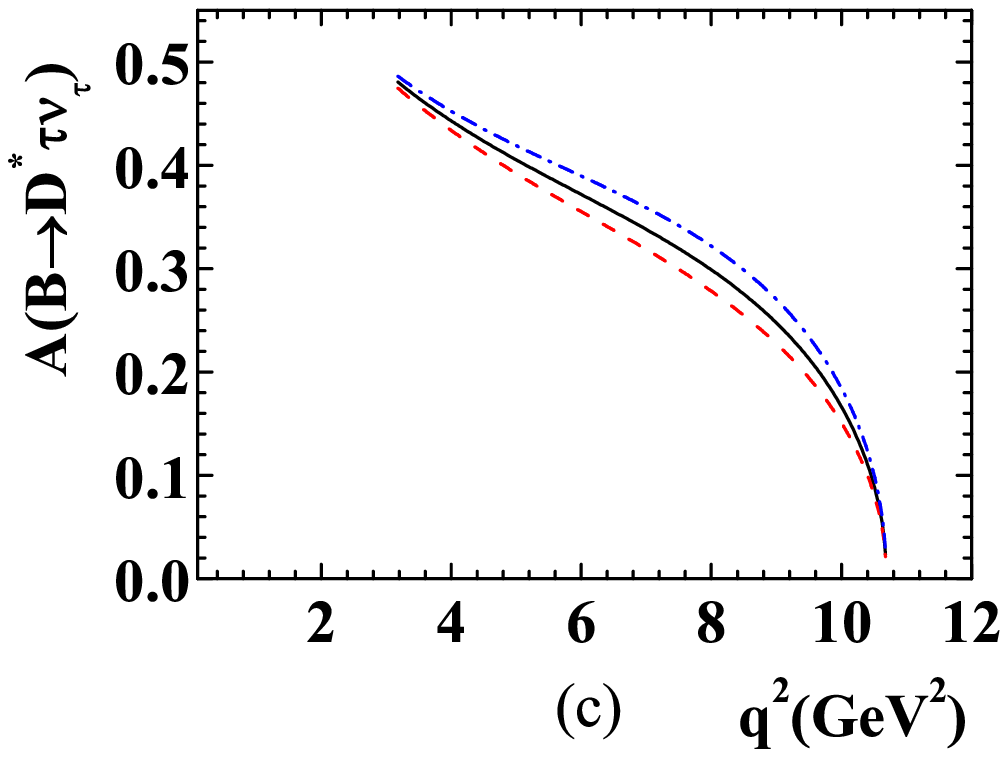} \caption{
Angular asymmetries with new effects for (a) $B\to D \tau
\bar{\nu}_{\tau}$, (b) $B\to D^* \ell \bar{\nu}_{\ell}$ and (c)
$B\to D^* \tau \bar{\nu}_{\tau}$. The solid lines denote the SM
results. The dashed (dash-dotted) lines  stand for (a)
$G_{S}=-0.22$ and (b,c) $G_{V(A)}=0.08\ (-0.08)$, respectively.}
 \label{figasy_new}
\end{figure}

Since $\tan\beta$ and the various masses are all free
parameters, in the following numerical estimations, we take
$\tan\beta=50$, $m_t=174$ GeV, $m_{b}=4.4$ GeV, $m_{H}=300$ GeV,
$m_{\tilde{g}}=|\mu|=A_{t}=A_{b}=200$ GeV, and $I_{0}=5$
\cite{WKN,WN}; and also, to maximize $C_{R}$, we set
$|\tilde{V}_{33}|=|V^{D_R}_{33}|=1$ and $|V^{U_R}|=1/\sqrt{2}$
\cite{WKN,WN}. Hence, we obtain $C^{\mu}_{H}=-0.01$,
$C^{\tau}_{H}=-0.22$ and $|C_{R}|\leq 0.08$.
%
%These diagrams have been studied in Refs. \cite{WKN,WN}, which
%gives  that $|G_S|\leq 0.35$ and $|G_V|=|G_A|\leq 0.08$. We note
%that the constraints on $G_{V,A}$ with a generic right-handed
%current have been studied in Ref. \cite{RC}, in which the allowed
%ranges of $G_{V,A}$ are much larger. To show the new physics
%effects, we take $G_S=\pm 0.2$ and $G_{A(V)}=\pm 0.08$.
The corresponding decay BRs due to the new physics are displayed
in Table \ref{tablebrnew}.
\begin{table}[phbt]
\caption{Decay BRs (in units of $10^{-2}$) of $B^-\to D^{(*)} \ell
\bar{\nu}_{\ell}$ with $G_{S}=-0.01$  and $-0.22$ for the $\mu$ and
 $\tau$ modes, respectively, and $G_{V,A}=0.08\ (-0.08)$.}\label{tablebrnew}
\begin{ruledtabular}
\begin{tabular}{ccccc}
 Decays & $B^-\to D^0 \mu \bar{\nu}_{\mu}$
  & $B^-\to D^0 \tau \bar{\nu}_{\tau}$ & $B^-\to D^* \ell \bar{\nu}_{\ell}$&
     $B^-\to D^* \tau \bar{\nu}_{\tau}$ \\
  \hline
 New Physics   & 2.07 & 0.84 & 6.25 (4.69) & 1.60 (1.20) \\
  \end{tabular}
\end{ruledtabular}
\end{table}
From the table, we see that they are consistent with
the experimental data \cite{PDG}.
% the constraints from BRs of $B\to D^{(*)} \ell \bar{\nu}_{\ell}$.
Our results for the angular
asymmetries  are
presented in Fig.~\ref{figasy_new}.
%
 %%%%%%%%%%%%%%%%%%%%%%%%%%%%%%%%%%%%%%%%%%%%%%%%%%%%%%%%%%%%%
 %%%%%%%%%%%%%%%%%%%%%%%%%%%%%%%%%%%%%%%%%%%%%%%%%%%%%%%%%%%%%
As seen from Fig. \ref{figasy_new},  the asymmetries in $B\to D^*
\ell \bar{\nu}_{\ell}$ with $\ell=e$ and $\mu$ are more sensitive to
the new physics from  the right-handed current in the quark
sector. Since we adopt the IW function to be
$\xi(\omega)=1+\rho^2 (\omega-1)$ \cite{GL,Babar}, by considering
the experimental errors for $\rho^2$, we find that the shapes for
angular distributions with different values of $\rho^2$ all
overlap each other, {\it i.e.}, our results of
Fig.~\ref{figasy_new} are insensitive to the errors from the $\rho^2$
parameter.

In summary, according to the lepton angular distributions  in
$B\to D^{(*)} \ell \nu_{\ell}$ decays, we have studied the angular
asymmetries with the general effective interactions. We have
illustrated the asymmetries in the quark currents with scalar and
$V+A$ interactions, respectively. We have shown that they are
sensitive to new physics with the right-handed quark current.\\

 \noindent {\bf Acknowledgments:}\\
This work is supported in part by the National Science Council of
R.O.C. under Grant \#s: NSC-91-2112-M-001-053, No.
NSC-93-2112-M-006-010 and No. NSC-93-2112-M-007-014.\\

%%%%%%%%%%%%%%%%%%%%%%%%%%%%%%%%%%%%%%%%%%%%%%%%%%%%%%
%\baselineskip=0.6cm


\begin{thebibliography}{99}
\bibitem{NP1}
P.~Krawczyk and S.~Pokorski,
%``Strongly Coupled Charged Scalar In B And T Decays,''
Phys.\ Rev.\ Lett.\  {\bf 60}, 182 (1988);
J.~Kalinowski,
%``Semileptonic Decays Of B Mesons Into Tau Tau-Neutrino In A Two Higgs Doublet
%Model,''
Phys.\ Lett.\ B {\bf 245}, 201 (1990);
B.~Grzadkowski and W.~S.~Hou,
%``Solutions to the B meson semileptonic branching ratio puzzle within two
%Higgs doublet models,''
Phys.\ Lett.\ B {\bf 272}, 383 (1991);
W.~S.~Hou,
%``Enhanced charged Higgs boson effects in B- $\to$ tau anti-neutrino, mu
%anti-neutrino and b $\to$ tau anti-neutrino + X,''
Phys.\ Rev.\ D {\bf 48}, 2342 (1993);
G.~Isidori,
%``Limits to the charged Higgs sector from B $\to$ X tau-neutrino,''
Phys.\ Lett.\ B {\bf 298}, 409 (1993);
Y.~Grossman,
%``Phenomenology of models with more than two Higgs doublets,''
Nucl.\ Phys.\ B {\bf 426}, 355 (1994);
Y.~Grossman and Z.~Ligeti,
%``The Inclusive anti-B $\to$ tau anti-neutrino X decay in two Higgs doublet
%models,''
Phys.\ Lett.\ B {\bf 332}, 373 (1994);
Y.~Grossman and Z.~Ligeti,
%``Transverse tau polarization in inclusive anti-B $\to$ tau anti-neutrino X
%decays,''
Phys.\ Lett.\ B {\bf 347}, 399 (1995);
Y.~Grossman, H.~E.~Haber and Y.~Nir,
%``QCD corrections to charged Higgs mediated b $\to$ c tau-neutrino decay,''
Phys.\ Lett.\ B {\bf 357}, 630 (1995);


\bibitem{NP2}
J.~G.~Korner, K.~Schilcher and Y.~L.~Wu,
%``T Odd And CP Odd Triple Momentum Correlations In The Exclusive Semileptonic
%Bottom Meson Decay B $\to$ D* Lepton Lepton-Neutrino,''
Phys.\ Lett.\ B {\bf 242}, 119 (1990);
C.~S.~Kim, J.~Lee and W.~Namgung,
%``CP violation in the semileptonic B(l4) (B $\to$ D pi l nu) decays:
%Multi-Higgs doublet model and scalar-leptoquark models,''
Phys.\ Rev.\ D {\bf 59}, 114006 (1999);
J.~P.~Lee,
%``CP violating transverse lepton polarization in B $\to$ D(*) l anti-nu
%including tensor interactions,''
Phys.\ Lett.\ B {\bf 526}, 61 (2002).

\bibitem{SC}
R.~Garisto,
%``CP violating polarizations in semileptonic heavy meson decays,''
Phys.\ Rev.\ D {\bf 51}, 1107 (1995);
M.~Tanaka,
%``Charged Higgs effects on exclusive semitauonic B decays,''
Z.\ Phys.\ C {\bf 67}, 321 (1995);
K.~Kiers and A.~Soni,
%``Improving constraints on tan(beta/m(H)) using B $\to$ D tau anti-nu,''
Phys.\ Rev.\ D {\bf 56}, 5786 (1997);
B.~Grzadkowski and W.~S.~Hou,
%``Searching for B $\to$ D tau anti-tau-neutrino at the 10-percent level,''
Phys.\ Lett.\ B {\bf 283}, 427 (1992).

\bibitem{RC} M.~B.~Voloshin,
%``Bound on V + A admixture in the b $\to$ c current from inclusive vs.
%exclusive semileptonic decays of B mesons,''
Mod.\ Phys.\ Lett.\ A {\bf 12}, 1823 (1997);
%[arXiv:hep-ph/9704278],
T.~G.~Rizzo,
%``Right-handed currents in B decay revisited,''
Phys.\ Rev.\ D {\bf 58}, 055009 (1998).
%[arXiv:hep-ph/9803385].

\bibitem{WKN}G.H. Wu, K. Kiers, and J.N. Ng, Phys. Lett. B{\bf
402}, 159 (1997);
G.H. Wu, K. Kiers, and J.N. Ng, Phys. Rev. D{\bf 56},
5413 (1997).

\bibitem{LQ}J.L. Hewett and T.G. Rizzo, Phys. Rev. D{\bf 56}, 5709
(1997); J.P. Lee, Phys. Lett. B{\bf 526}, 61 (2002).

\bibitem{GL} B. Grinstein and Z. Ligeti, Phys. Lett. B{\bf 526}, 345
(2002).

\bibitem{TP} J.~G.~Korner and G.~A.~Schuler,
%``Exclusive Semileptonic Decays Of Bottom Mesons In The Spectator Quark
%Model,''
Z.\ Phys.\ C {\bf 38}, 511 (1988)
[Erratum-ibid.\ C {\bf 41}, 690 (1989)].


\bibitem{PDG} Particle Data Group, S. Eidelman {\it et al.}, Phys. Lett. B{\bf 592}, 1
(2004).

\bibitem{WN}G.H. Wu and J.N. Ng, Phys. Rev. D{\bf 55}, 2806
(1997).

\bibitem{Babar}BABAR Collaborations, B. Aubert {\it et al.},
hep-ex/0409047.


\end{thebibliography}
\end{document}